# Rare frustration of optical supercontinuum generation

D. R. Solli[1], C. Ropers[1,2], B. Jalali[1]

[1]Department of Electrical Engineering, University of California, Los Angeles, 90095 USA

[2]Courant Research Center Nano-Spectroscopy and X-Ray Imaging, University of Göttingen, Germany

**Extremely large, rare events arise in various systems, often representing a defining character of their behavior[1,2,3,4,5,6,7,8]. Another class of extreme occurrences, unexpected failures, may appear less important, but in applications demanding stringent reliability, the rare absence of an intended effect can be significant. Here, we report the observation of rare gaps in supercontinuum pulse trains, events we term *rogue voids*. These pulses of unusually small spectral bandwidth follow a reverse-heavy-tailed statistical form. Previous analysis has shown that rogue waves, the opposite extremes in supercontinuum generation, arise by stochastic enhancement of nonlinearity[9,10]. In contrast, rogue voids appear when spectral broadening is suppressed by competition between pre-solitonic features within the modulation-instability band. This suppression effect can also be externally induced with a weak control pulse.**

Many dynamic, complex systems in natural, industrial, and societal contexts produce rare, extreme events or outliers—rare instances of strongly atypical behavior. Although infrequent and short-lived, such anomalous events can have a defining influence on the long-term condition of the system. Stock market crashes, natural disasters, pandemics, and freak ocean waves testify to the existence and impact of these unusual events. Heavy-tailed distributions, the signature of such phenomena, assign much more substantial probabilities to anomalous events than normal distributions[1,2].

Another class of extreme occurrences often goes unnoticed, but deserves equal attention: sudden and unexpected failures, singular absences of otherwise repetitive or standard behavior, or simply a lone event of extremely small magnitude. In systems and applications that require reliability of a known cause-effect (or input-output) relation, such anomalies can be critical. Examples of systems known to display extreme value behavior and in which unexpected failures could have significant consequences can be found in very diverse areas such as finance[3], mechanical reliability[4], networking[5], ecological diversity[6], genetic expression[7], and unexploded ordnance[8]. These unusual events—observed when input fluctuations are mapped into skewed output statistics by a nonlinear relationship—may also signify important physical characteristics of the underlying complex system that are otherwise difficult to identify.

Interestingly, extreme events also appear in the realm of ultrafast phenomena. It has recently been shown, for example, that extremely rare white light pulses, optical rogue waves, can arise in nonlinear optical experiments[9,10]. In other optical contexts, non-Gaussian statistics can arise from gain fluctuations and in soliton-based communication systems[11,12]. Optical experiments offer a particularly convenient platform to study rare phenomena because many of the requisite experimental variables can be readily manipulated. In addition, optical timescales permit a statistically significant number of extreme events to be captured in a brief period, even though they may be rare.

Here, we report the direct observation of optical *rogue voids* (RVs), anomalous events corresponding to a rare suppression of nonlinear spectral broadening during the generation of broadband radiation known as supercontinuum (SC). These events follow a left-skewed or reverse-heavy-tailed distribution in which events far below the mean occur with small but non-negligible probability. Given sufficient input power, it is expected that each member of a nearly-identical population of input pulses will be converted to an ultra-broadband continuum by nonlinear action. Pulse-resolved, real-time measurements reveal that a small fraction of pulses experiences significantly less broadening, leading to rare voids in the SC pulse train (cf. Figure 1A). As explained in more detail below, a stochastic anti-seeding effect causes the SC generation process to be slowed on occasion.

Although optical experiments offer a convenient platform for the study of extreme phenomena, challenges do arise in the detection of non-repetitive, ultrashort events. Using ultrafast real-time analog-to-digital converters (ADCs), extreme spectra can be detected in the fast pulse series supplied by a mode-locked laser (see Methods Summary). With this method, we are able to detect the presence of singular events during SC generation. To produce SC, we inject pump pulses from a mode-locked laser into a segment of highly nonlinear optical fiber. At the fiber output, a spectral region redshifted from the input wavelength is selected with a bandpass filter. By measuring the time-averaged power passed by the filter, we determine the input power level needed to generate significant spectral broadening (cf. Figure 1B). The energy passed by the filter shows a threshold response to input power, rising sharply initially and saturating for larger input power levels. Above threshold, a time-averaged spectral measurement of the output SC records a broadband spectrum covering the filter's passband. Next, we perform real-time detection of the filtered signal. While rogue waves appear below threshold, the regime above threshold produces a different extreme-value distribution in which anomalous events—pulses that experienced much less broadening than expected—appear as rare gaps in the SC pulse train (cf. Figures 1C and 1D). Time-averaged power measurements do not detect these events.

We also model this effect by solving the nonlinear Schrödinger equation (NLSE), which is widely used to study SC generation in nonlinear fiber[13]. We include weak broadband stochastic perturbations in the temporal envelope of the Gaussian input field and simulate propagation in the nonlinear fiber for many independent events (see Methods Summary). Above the input power threshold, we again find a reverse heavy-tailed distribution for the redshifted energy (cf. Figure 2A), in qualitative agreement with our experimental observation. Events that are more than four orders of magnitude and 10 standard deviations below the mean appear within the extended tail. These events have broadened much less than expected (cf. Figure 2B).

In SC generation, large spectral broadening factors are obtained when intense narrowband pulses are injected into a nonlinear fiber at its wavelength of zero dispersion. In the initial stages, modulation instability (MI)[14] amplifies certain components of the ambient noise, creating spectral sidebands that flank the input spectrum. Once these modulation sidebands gain sufficient amplitude, pulse breakup begins. In this process, the input light spontaneously fissions into solitons plus blueshifted dispersive radiation[15]. With continuing propagation, the Raman self-frequency shift causes the solitons to progressively shift towards longer wavelengths. Soliton fission and the Raman self-frequency shift transform the input field into a broadband waveform with significant redshifted energy. Since energy from a gain process (MI) must build to a critical level to trigger this rapid broadening, the redshifted energy displays a threshold response to input power. As the gain is seeded by noise, the pulse-to-pulse amplitude stability and phase coherence are lost[13,16,17,18,19].

In this process, the input pulse power and noise level are both important parameters. However, the mechanism is specifically sensitive only to noise with frequency content and timing capable of seeding MI[9,10]; below threshold, a random surplus in this noise component can produce a rare redshifted soliton or rogue wave. This process can also be exploited to influence SC generation with an external signal[20,21,22] or feedback loop[23]; stimulating the process with a controlled signal, for example, results in a much more stable source of white light[20,21].

RVs are observed when the input power and noise level should be sufficient to produce a broadband output spectrum. To explore the mechanism behind these rare occurrences, we return to the NLSE-based simulations. Interestingly, an input-output comparison reveals that RVs are not generally correlated with comparatively low-amplitude noise components. This finding indicates that these anomalous events are not a simple corollary of the rogue wave phenomenon. The energy contained within the MI sidebands is comparable for RVs and normal events prior to the typical onset of soliton fission. Yet for RVs, broadening is suppressed before soliton fission should begin.

To study this mechanism in greater detail, we systematically remove different noise components from the initial conditions (ICs) with a scanning time-frequency filter. By selectively removing specific components of the noise, we can determine each component's impact on the broadening process. After removing only a specific noise component, we calculate the redshifted energy generated by the modified IC after passage through the nonlinear fiber. Performing the procedure for a raster of times and frequencies generates a map relating the noise to the SC bandwidth. The noise-filtering process reveals a notable feature: for the RVs, typical redshifted energy would have been produced if a small component of the input noise had been *absent* (cf. Figures 3A and 3B). Furthermore, a redshifted soliton becomes clearly visible in the output time-frequency profile when this noise component is deleted from the IC (cf. Figures 3C and 3D). This observation suggests that a component of the noise can actively suppress or delay spectral broadening.

Motivated by these observations, we investigate the possibility of such an anti-seeding effect in SC generation. As mentioned above, previous experiments have shown that a weak coherent seed pulse can be used to stimulate SC generation instead of input noise[20]. Beginning with initial conditions suitable for coherent stimulated SC generation, we add an additional weak pulse. Varying the relative frequency shift and timing of this additional pulse, we calculate the new SC in each case to search for a possible anti-seeding condition. In the absence of this pulse, a broadband SC is generated as expected. But when added with the correct timing and frequency shift, the anti-seed suppresses SC generation, producing the stimulated form of a rogue void, with nearly identical spectral and temporal characteristics (cf. Figure 4).

The normal seeding process stimulates SC generation by providing a small amplitude modulation on the IC; this modulation is amplified by MI, and evolves into a soliton. Induced MI has been previously explored for a variety of applications[14]. For example, it has been used to generate high-repetition rate pulse trains[24], promote Raman soliton formation[25], and enhance and stabilize SC generation[20] as discussed. Here, we observe that a very weak signal can also delay SC generation by influencing MI. Addition of the anti-seed increases the temporal span of the modulation on the initial pump envelope. Close examination of this new modulation reveals that it evolves into two independent sharp features rather than just one. Soliton fission is delayed because these features compete for MI gain by depleting the local region of the pump envelope. Thus, the MI energy is spread out between these features and not into a single pre-solitonic wavepacket. While it may be expected that two such features can deplete the pump, it is not obvious that the additional modulation supplied by the anti-seed—also contained within the MI band—can result in such a significant drop in the SC bandwidth.

Inspection of the nonlinear evolution of the RVs shows similar but stochastically-initiated dynamics, illustrated in Figures 5A and 5B. Generally, multiple pre-solitonic features form in the RV events (evidenced by RV temporal profiles) and the MI energy is spread between them. In contrast, the normal events usually contain one dominant sharp waveform, which develops into a redshifted soliton. Additional features do not substantially delay this process if they do not deplete the same portion of the pump envelope or if they coalesce at an early stage. Indeed, an analysis of a large population of events shows that RVs tend to have a greater number of sharp wavepackets than normal events (cf. Figure 5C). With sufficient propagation in a low-loss fiber, a given RV would eventually produce multiple redshifted solitons, but this does not occur within the specified length. These rare events provided the initial clue that the broadening process can also be intentionally delayed with a specialized weak field.

In conclusion, we have observed rare frustrated events in optical SC generation. These rogue voids arise due to competition between pre-solitonic features seeded by the input noise. Deleting a small portion of the input noise can interrupt this competition and restore normal SC generation. Using this process, supercontinuum generation can be intentionally suppressed by a weak control signal. This effect creates numerous new possibilities for switching and coherent control of nonlinear broadening.

Acknowledgments: This work was performed under the PHOBIAC program supported by the Defense Advanced Research Project Agency (DARPA).
C. R. acknowledges funding by the German Initiative of Excellence (FL3).

Methods Summary

Conventional spectrometers, although able to operate in single-shot mode, are not suited to capture a large number of ultrafast events in the search for rare extremes. A new experimental technique, inspired by time-stretch analog-to-digital conversion[26], which has been used to detect optical rogue waves, measure Raman spectra in real time[27], and acquire optical images at unprecedented frame rates[28], employs group-velocity dispersion to transform a real-time oscilloscope into an ultrafast spectrometer. This wavelength-time transformation permits real-time detection of multiple sample points within a single ultrashort rogue event, and provides spectral information. As such, it was instrumental in providing a clear identification of optical rogue waves when they were initially reported[9]. However, with this knowledge in hand, full spectral characterization is not necessarily needed to detect events with unusual spectral content: extreme spectra can be identified by the energy in their redshifted spectral tails[9,10,20,29] because integration of the spectral tail will also reveal the presence of an unusually broadband event. This parameter can be measured using a suitably chosen wavelength filter (designed to pass only wavelengths far from the pump), a photodetector, and a real-time ADC.

In our experiments, we employ an input stream of near transform-limited pulses centered at 1550 nm (duration ~3 ps) with a repetition rate of 25 MHz from an amplified mode-locked laser. To produce SC, these pulses are delivered to a segment of highly nonlinear fiber ($L = 15\,\text{m}$). The fiber's relevant parameters are $\beta_2 = 1.13 \times 10^{-4}\,\text{ps}^2/\text{m}$, $\beta_3 = 6.48 \times 10^{-5}\,\text{ps}^3/\text{m}$, $\gamma = 10.66\,\text{W}^{-1}\text{km}^{-1}$ (nonlinear coeff.), and $T_R \approx 5\,\text{fs}$ (Raman response parameter). At the output, a bandpass filter ($\lambda_0 = 1705$ nm, $\Delta\lambda = 48$ nm) selects a spectral region well removed from the input wavelength.

In our stochastic simulations, we include complex noise in the temporal envelope of the Gaussian input field according to the prescription:

$$A(z=0,t) = \sqrt{P_0}\, e^{-(t-t_0)^2/\tau^2}[1 + \varepsilon(r_1(t) + i \times r_2(t))],$$

where $\tau$ defines the pulse duration, $P_0$ is the input peak power, $\varepsilon$ is the noise coefficient, and $r_1(t)$ and $r_2(t)$ represent random numbers selected at each point in time from a normal distribution with zero mean and unit variance. A noise coefficient of $\sim 4 \times 10^{-5}$ is employed along with input peak power of 210 W. In order to impose a reasonable limit on the noise bandwidth, we apply a Fourier filter with a 30 THz bandwidth about the input wavelength. In contrast to simple input power fluctuations, this noise produces weak broadband fluctuations across the pulse envelope. We apply the NLSE to a given IC and spectrally filter the output to determine the amount of redshifted energy (>1870 nm). To produce stimulated SC we add a very weak seed pulse with 200 fs duration at a delay and frequency shift of -2 ps and -10 THz relative to a noiseless pump. Without seeding, very little redshifted energy is produced by the pump. Optimal anti-seeding is achieved by adding another pulse identical to the seed, but with a slightly different delay and frequency shift (~ -2.2 ps and -9.9 THz). Other anti-seed parameters are also possible.

Figure 1

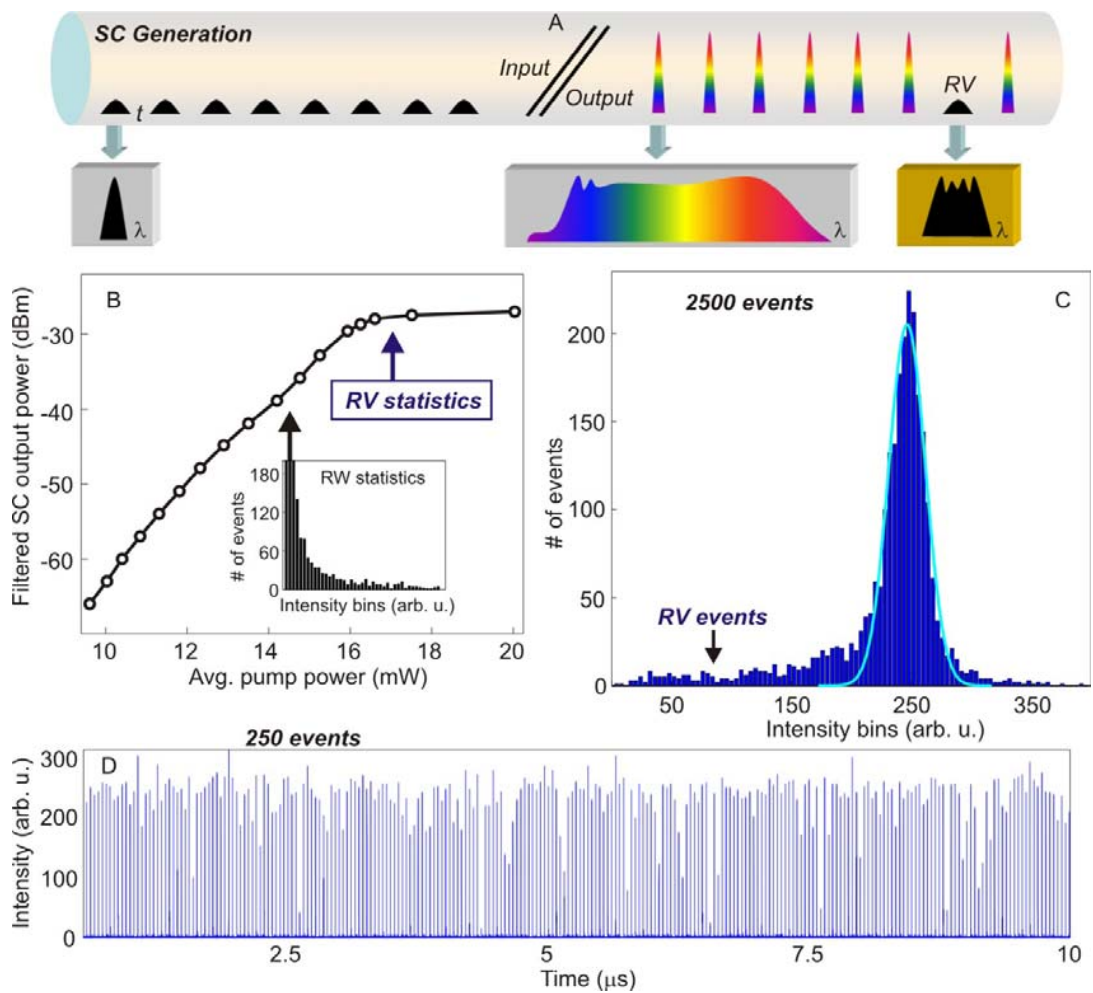

**Figure 1 Experimental observation of rogue voids.** A) Conceptual illustration of a rogue void (RV) in supercontinuum (SC) generation. Narrowband input pulses injected into nonlinear fiber produce a SC pulse train with occasional misfires. B) Experimental measurement of red-filtered SC output power vs. average power of input pulse train. The nonlinear pulse transformation displays a threshold behavior. Rogue wave (RW) statistics are observed below threshold, RV statistics appear above threshold. C) Experimental observation of reverse-heavy-tailed statistics at the input power level labeled by the arrow in B). A Gaussian curve is fit to the main portion of the distribution, highlighting the RVs—events that are very improbable in normal statistics. D) Filtered SC pulse train measured in real time: RVs are observed.

Figure 2

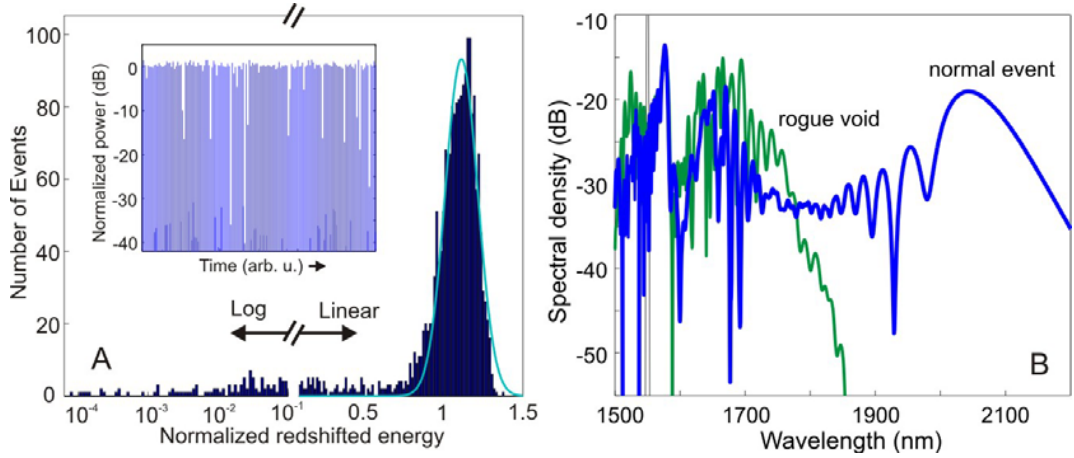

**Figure 2 Rogue voids appear in simulations.** A) Reverse-heavy-tailed distribution calculated from 2000 independent events with input pulse power of 210 W and weak broadband input noise as described in Methods Summary. Event magnitude is calculated by integrating the redshifted energy (>1870 nm). A Gaussian curve is fit to the main portion of the distribution; a portion of the extreme tail is shown in log scale. Inset: a time sequence of ~250 redpass filtered output SC pulses (normalized power >1870 nm, dB scale). B) Spectra of one rogue void (green) and one normal event (blue). Pump (input) spectrum is also shown (black).

Figure 3

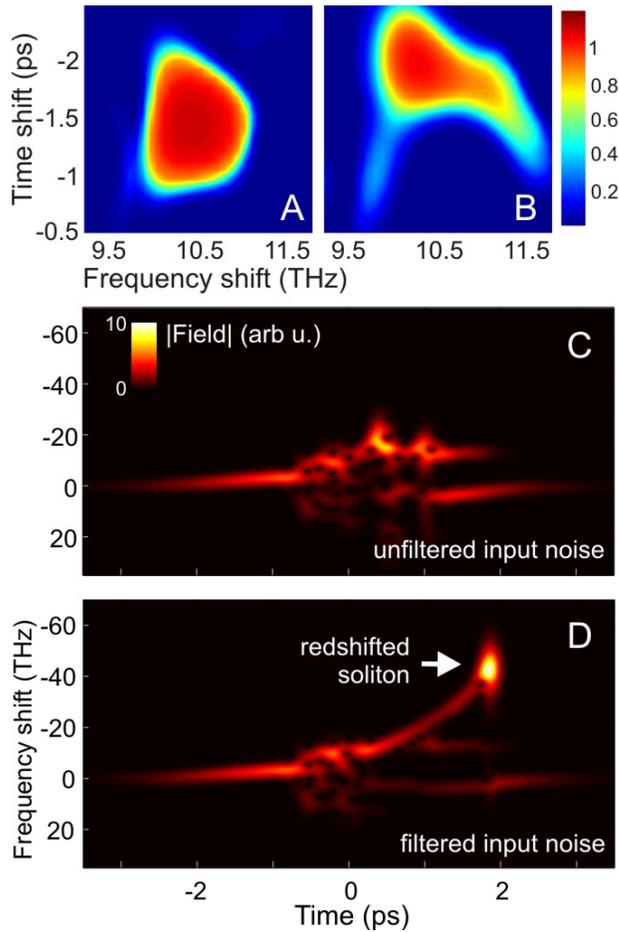

**Figure 3 Deconstructing rogue voids.** A scanning time-frequency filter (Gaussian profile, $\Delta t = 1\,\text{ps}$; $\Delta f = 1\,\text{THz}$; center position specified by x and y axes) is applied to the input noise profiles of two different rogue voids (A, B). The filter removes only the noise components within its stop band. The plots are generated by sweeping the filter through the input noise and repeating the nonlinear simulation for each filter position. The process shows how each component of input noise impacts the output SC. The color scale represents the redshifted energy (>1870 nm) at the fiber output for each filter position. Removing a key noise component strongly increases the redshifted energy. Short-time Fourier transforms (field magnitude) of a rogue void before (C) and after (D) a key noise component is removed. After removing this tiny component, the IC produces an intense redshifted soliton.

Figure 4

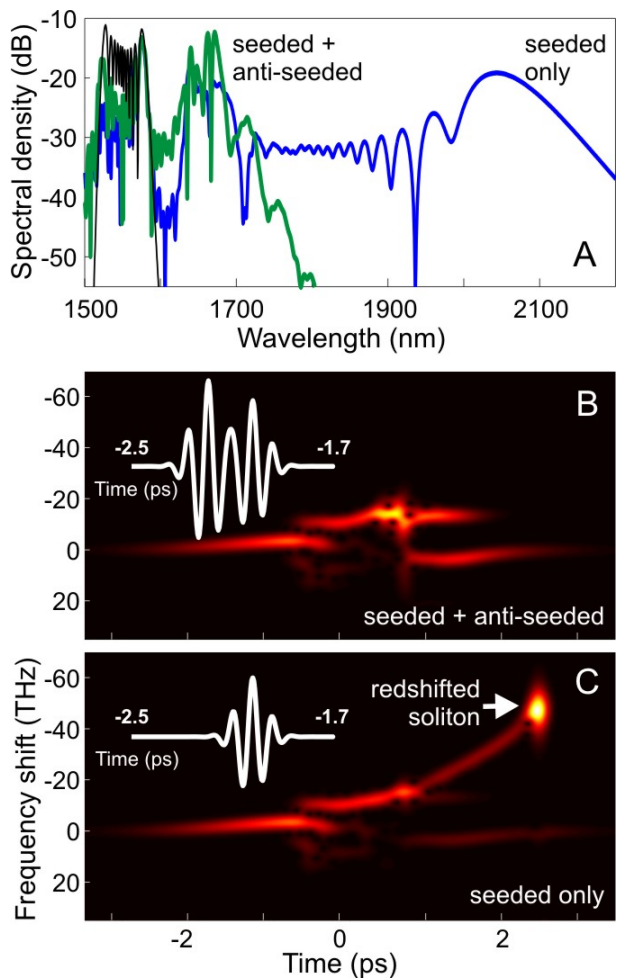

**Figure 4 Anti-seeding of supercontinuum generation.** Instead of input noise, a weak seed pulse with specific frequency shift and timing accelerates supercontinuum (SC) generation by producing a small intensity modulation of the pump envelope. Adding a second weak pulse (anti-seed) with slightly different timing and frequency shift can strongly suppress the seeding effect. A) Spectra of the seeded SC (blue) and the anti-seeded SC (green). Unseeded SC spectrum (black) also shown. Seed and anti-seed parameters are described in Methods Summary. B, C) Short-time Fourier transforms (field magnitude) of the anti-seeded and seeded SC, respectively. Insets show the tiny input intensity modulation of the pump envelope with (top) and without (bottom) the anti-seed. The spectral and temporal profiles of the anti-seeded SC are similar to those of a rogue void.

Figure 5

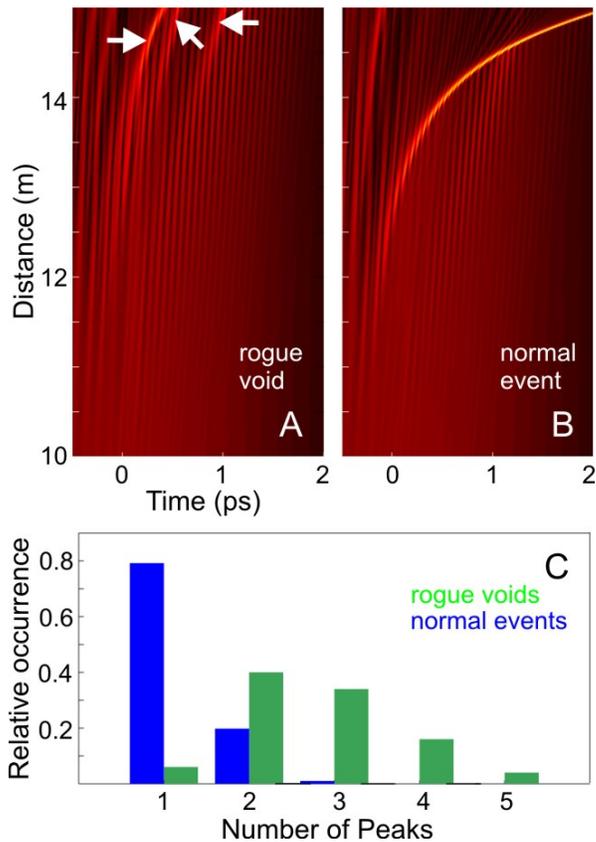

**Figure 5 Rogue void evolution.** Propagation of A) a rogue void (RV) and B) a normal event. Color scale illustrates the field magnitude as a function of time (x-axis) and distance propagated in the fiber (y-axis). Bright streaks track the evolution of a sharp pulse in time. The RV contains three competing pre-solitonic features of similar amplitudes. The normal event develops a single dominant feature; with no significant competition, it rapidly evolves into a redshifted soliton (delay increases with propagation due to Raman scattering). C) Statistical analysis of the number of separate peaks above an arbitrary threshold in normal events (blue) and RVs (green) at the end of the fiber. Rogue voids tend to contain more distinct peaks, which compete for MI gain. Some normal events may contain more than one peak: if sufficiently separated in time and/or space, the competition is minimal.